%% file: main.tex
\documentclass[runningheads]{llncs}
\usepackage{graphicx}
\usepackage[utf8]{inputenc}
\usepackage{lipsum}
\usepackage{float}
\usepackage{abstract}
\usepackage{amsmath}
\usepackage{hyperref}
\usepackage{xurl}
\usepackage[noabbrev,capitalize]{cleveref}
\usepackage{adjustbox}
\usepackage[sort&compress,numbers]{natbib}

\usepackage{fancyhdr}
\newcommand\blfootnote[1]{%
  \begingroup
  \renewcommand\thefootnote{}\footnote{#1}%
  \addtocounter{footnote}{-1}%
  \endgroup
}

\title{Teacher-Student Architecture for Mixed Supervised Lung Tumor Segmentation}

\author{Vemund Fredriksen\inst{1,\ddag} \and %\orcidID{0000-0002-3887-7399} \and
Svein Ole M. Sevle\inst{1,\ddag} \and %\orcidID{0000-0002-5581-3629} \and
Andr\'e Pedersen\inst{2,3,4,*} \and %\orcidID{0000-0002-3637-953X} \and
Thomas Langø\inst{2,5} \and %\orcidID{0000-0002-2824-6120} \and
Gabriel Kiss\inst{1,5} \and % \orcidID{1111-2222-3333-4444} \and
Frank Lindseth\inst{1} % \orcidID{0000-0002-4979-9218}
}

\authorrunning{ }
\titlerunning{ }

\institute{Department of Computer Science, Norwegian University of Science and Technology (NTNU), Trondheim, Norway \and
Department of Health Research, Medical Technology, SINTEF, Trondheim, Norway \and
Department of Clinical and Molecular Medicine, Norwegian University of Science and Technology (NTNU), Trondheim, Norway \and
Clinic of Surgery, St. Olavs hospital, Trondheim University Hospital, Trondheim, Norway \and
Research Department, Future Operating Room, St. Olavs hospital, Trondheim University Hospital, Trondheim, Norway \\
}

\date{August 2021}

\begin{document}

\maketitle

\input{Sections/01_abstract}

%\thispagestyle{firstpage}

%\newpage
%\thispagestyle{otherpages}
\input{Sections/02_introduction}
\input{Sections/03_materials-and-methods}

\input{Sections/05_results}
\input{Sections/06_discussion}
\input{Sections/07_conclusion}
\input{Sections/08_acknowledgements}
\input{Sections/09_declarations}

\bibliographystyle{splncs} %{splncs03} %{unsrtnat} %{unsrt}

\section*{References}
\bibliography{references.bib}
%\printbibliography[heading=none]{references.bib}

\end{document}

%% file: Sections/01_abstract.tex
\begin{abstract}
    \noindent\textbf{Purpose} Automating tasks such as lung tumor localization and segmentation in radiological images can free valuable time for radiologists and other clinical personnel.
    Convolutional neural networks may be suited for such tasks, but require substantial amounts of labeled data to train. Obtaining labeled data is a challenge, especially in the medical domain.
    \blfootnote{Preprint submitted December 21, 2021}
    \blfootnote{\ddag \vspace{1pt} These authors contributed equally to this work.}
    \blfootnote{* Corresponding author: andre.pedersen@sintef.no}
    \newline
    \textbf{Methods} This paper investigates the use of a teacher-student design to utilize datasets with different types of supervision to train an automatic model performing pulmonary tumor segmentation on computed tomography images. The framework consists of two models: the student that performs end-to-end automatic tumor segmentation and the teacher that supplies the student additional pseudo-annotated data during training. 
    \newline
    \textbf{Results} Using only a small proportion of semantically labeled data and a large number of bounding box annotated data, we achieved competitive performance using a teacher-student design. Models trained on larger amounts of semantic annotations did not perform better than those trained on teacher-annotated data.
    \newline
    \textbf{Conclusions} Our results demonstrate the potential of utilizing teacher-student designs to reduce the annotation load, as less supervised annotation schemes may be performed, without any real degradation in segmentation accuracy.
    \keywords{Teacher-Student Framework \and Pulmonary Tumor Segmentation \and Deep Learning \and Convolutional Neural Networks \and Computed Tomography}
\end{abstract}

%% file: Sections/02_introduction.tex
\section{Introduction}
Cancer is becoming the leading cause of death and the most significant obstacle to increase life expectancy in many countries~\cite{world2020report}. Lung cancer, accounting for more than 11~\% of all new cases, is the second most common cancer and it ranks first among the cancer-related mortality worldwide, accounting for 18~\% of the total cancer deaths~\cite{Sung2021}.
The most common lung cancer treatments include: surgical resection, chemotherapy, radiotherapy, and immunotherapy. Many of these treatments, and also the successful diagnosis with bronchoscopy or computed tomography (CT)-guided biopsy, depend on accurately locating, and in many cases delineating (segmenting), the tumor from normal tissue in the preoperative images, typically CT. 

Manual segmentation of the lesions/tumors from preoperative CT is a laborious and tedious process for oncologists, radiologists, and pulmonologists, which could result in delays of treatment and lower the survival rates, especially in clinics with inadequate resources.
In addition, the quality of manual localization and segmentation relies on prior knowledge and clinical expertise. Even with adequate guidelines and standards, tumor segmentation is often prone to high inter- and intraobserver variability.
%inconsistencies may still exist in the segmentations, both with respect to inter and intra-observer variability. 
On the other hand, automatic segmentation techniques has the potential to provide efficient, consistent, and more accurate results. Automatic methods can both shorten the time needed to read the images and they also allow experts to devote their limited time to optimize planning and treatment planning.

Historically, methods like thresholding~\cite{UZELALTINBULAT2017}, region growing~\cite{Adams1994,Dehmeshki2008}, and graph cuts~\cite{GU2013692} were commonly proposed to segment lung tumors from CT images. These algorithms are suitable as semi-automatic methods, but are not suited for localization of lung tumors.
Recent advancements in deep learning enables automation of tasks that until recently was only performed by trained experts~\cite{ramanto2019,kim2019}. Advancements in hardware has enabled development of larger and increasingly complex models, but much of the improvement is caused by access to large amounts of annotated data. 
 
Today, especially after the introduction of the U-Net~\cite{ronneberger2015} architecture, deep learning methods have dominated the field of medical image segmentation~\cite{Du2020}. However, convolutional neural networks (CNNs) are memory intensive, especially for 3D volumes. It is therefore common to train networks based on 2D or 2.5D input images~\cite{carvalho}, where the model evaluates one slice at a time, chunks of slices, or 3D patches, and then applied in a sliding windows fashion across the CT volume. Patching the 3D volume comes at a cost of loss of perception, and thus more efficient multi-scale 3D CNN architectures have been proposed, which enables the use of larger input volumes~\cite{Jiang2018}. An alternative approach is to perform segmentation in multiple steps, either using multiple algorithms or a cascade of CNNs~\cite{Hossain2019,isensee,carvalho,gan2021}.

To accommodate the issue of lacking training data, unsupervised methods like supervoxel has been proposed~\cite{Hansen2021}. To facilitate faster convergence and more accurate results, multi-modality methods that utilize magnetic resonance imaging (MRI) or positron emission tomography (PET) scans in addition to CT have been suggested~\cite{fu2021,Jiang2019,Jiang2019_2}. Neural network architectures that can utilize multiple annotation types has also been suggested~\cite{Wang2019,Mlynarski2019}.

A more recent strategy to accommodate the lack of training data is the teacher-student design, inspired by the concept of knowledge distillation~\cite{phuong19,furlanello18,hinton2015}. The teacher creates pseudo-annotations from suboptimal annotations to increase the dataset size for training the student. The teacher-student pattern can be applied to any type of network architecture, and does not dictate other hyperparameters or external configurations. A teacher-student design can be used in different ways, from utilization of unlabeled data~\cite{caron2021,Xie2020,Li2020}, to exploitation of multiple modalities~\cite{Jiang2019,Jiang2019_2,Li2020}, and to usage of datasets with different annotation types~\cite{sun2020,Zhang2021}.

Our approach differs from the previously mentioned methods applied to lung tumor segmentation by utilizing annotations of different supervision on the same modality, namely CT images. Since CT scanning is less invasive to the patient than the other modalities, it is a goal to efficiently segment tumors from CT-only examinations. Our method is inspired by Sun \textit{et al.}~\cite{sun2020} that shows promising results using a teacher-student framework to segment liver and liver lesions given semantic and bounding box annotations. To the best of our knowledge, we are the first to implement a similar teacher-student framework for CT images to perform semantic segmentation of lung tumors. Our study suggests that even with a small dataset of semantic annotations, a student can achieve state-of-the-art performance given a large enough pseudo-annotated dataset to learn from.

%% file: Sections/03_materials-and-methods.tex
\section{Materials and methods}
\begin{figure}[h!]
    \centering
    \includegraphics[width=1.0\textwidth]{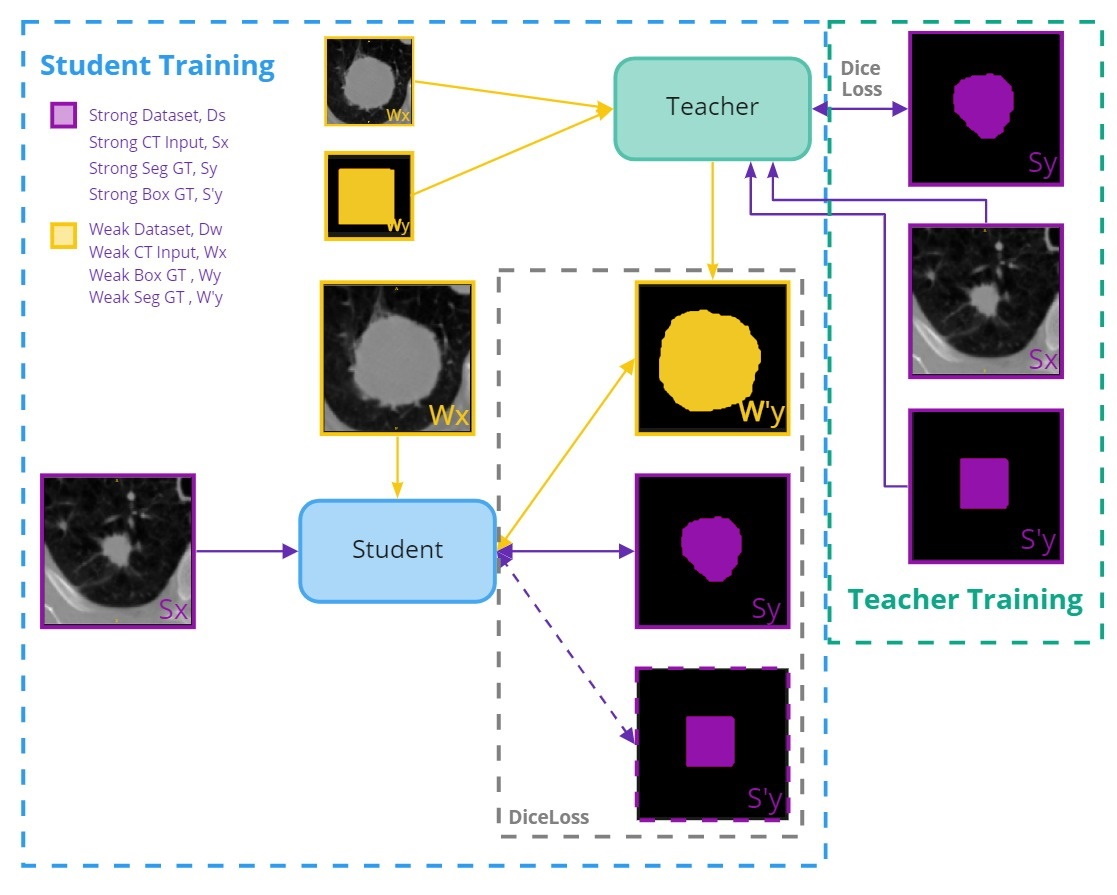}
    \caption{The method overview during training. Firstly, the teacher was trained using the strong dataset (semantic annotated images) represented with purple lines. The teacher was then used to make semantic annotations for the weak dataset (bounding box annotated images). The student was then trained on both the pseudo-annotated dataset $D_{w'}$, represented by the orange lines, and the strong dataset $D_s$. We trained two different students, only one of them were backpropagated using the box annotations, indicated by the dashed line.}
    \label{fig:method_overview}
\end{figure}

Our method consists of two separate models: a semi-supervised teacher and a fully-automatic student. The method relies on two different annotation types: semantic 3D annotations and 2D bounding boxes in the axial planes. These we refer to as \textit{strong} and \textit{weak} annotations. Furthermore, we define our strongly and weakly annotated datasets as $D_s = {\{(X_{i}^{s}, y_{i}^s)\}}_{i=1}^{m_{s}}$ and $D_w = {\{(X_{i}^{w}, y_{i}^w)\}}_{i=1}^{m_{w}}$, respectively. An overview of our design can be seen in Fig. \ref{fig:method_overview}.

%It is crucial that there exists an algorithm to generate a weakly labeled dataset from the strongly labeled set $D_{s'} = {\{(X_{i}^{s}, y_{i}^{s'})\}}_{i=1}^{m_{s}}$. It is straight-forward to compute the bounding boxes in each plane given the 3D segmentation of the tumor. The teacher is trained on $\{D_s, D_{s'}\}$ to produce semantic 3D segmentation, $D_{w'}$, given CT image and bounding box annotations as input.

\subsection{Data}
To study the effect of our teacher-student framework we used three public datasets: Medical Segmentation Decathlon (MSD)-Lung~\cite{simpson2019}, Non-Small Cell Lung Cancer (NSCLC)-Radiomics~\cite{radiomics_publication,radiomics_dataset}, and Lung-PET-CT-Dx~\cite{lungdx_dataset}. All three datasets contain manual annotations by human experts. The first two datasets consist of semantic annotations, whereas the latter dataset contains bounding box labels annotated in the axial plane.

The MSD-Lung dataset contains 64 images, whereas the NSCLC-Radiomics and Lung-PET-CT-Dx datasets contains 422 and 1295 images, respectively. Multiple images in the Lung-PET-CT-Dx dataset were discarded. The discarded images were either PET or PET/CT-fused, only contained a small portion of the thorax, or comprised of multiple scans stacked on top of each other. After removing all non-CT images and images with a real-world length (Z-axis) outside the range $[16, 60]$~cm, 665 images from Lung-PET-CT-Dx remained in our dataset.

\begin{table}[H]
    \centering
    \caption{Tumor sizes of the three datasets}
    \begin{tabular}{l|cc}
        \textbf{Dataset} & \textbf{Volume [$\boldsymbol{cm^3}]$} & \textbf{Diameter [$\boldsymbol{mm}]$} \\
        \hline
        MSD-Lung & $21.98 \pm 51.66$ & $37.63 \pm 20.08$\\
        NSCLC-Radiomics & $75.37 \pm 96.30$ & $63.63 \pm 29.62$\\
        Lung-PET-CT-Dx & $63.67 \pm 86.26$ & $48.66 \pm 19.85$\\
    \end{tabular}
    \label{tab:tumor_size_datasets}
\end{table}

The three datasets varied in terms of voxel density and tumor sizes. Overall, the Lung-PET-CT-Dx and the NSCLC-Radiomics datasets contain larger tumors than the MSD-Lung dataset (see Table \ref{tab:tumor_size_datasets}).
%Table \ref{tab:tumor_size_datasets} presents the average tumor size measured in volume and diameter. 
The tumor diameter is an approximate size, measured by calculating the average of the longest and shortest diameter of the tumor in real-world coordinate space.

\subsection{Preprocessing}
Our preprocessing pipeline consisted of multiple steps. Firstly, the voxel intensities were clipped to the range [-1024, 1000], before being standardized using the Z-score normalization method. The images' voxel spacing were then normalized to an anisotropic resolution of $1\times1\times1.5$ mm$^3$. Lastly, a volumetric cropping was applied, which differed between the teacher and the student.

For the teacher, the images were cropped around the tumor with a fixed resolution of $128\times128\times128$ voxels, whereas for the student, the images were split in two, each cropped around one of the lungs. The lungs were automatically segmented using the \texttt{lungmask} command line tool~\cite{hofmanninger2020}, and used when performing cropping around the lungs. The ground truth label images were voxel normalized and cropped in a similar manner as their corresponding CT image.

\subsection{Teacher-Student design}
The teacher was trained on 3D patches surrounding the tumor, guided by the corresponding bounding box annotations.
Once trained, the teacher was applied to $D_w$ to generate pseudo-strong labels, $D_{w'}$. Although expert labeled images are the gold standard, teacher pseudo-annotated images can enhance training of fully automatic models, or even be used to aid experts in clinical use.

The student, like any ordinary automatic method, takes CT images as input and produces 3D segmentations of the potential lung tumors without user interaction. During training, the student exploits the pseudo-annotated images in $D_{w'}$ produced by the teacher, using the extended dataset, $\{D_{s}, D_{w'}\}$.
%Given that the extended dataset, $\{D_{s}, D_{w'}\}$, is vast enough compared to the strong data set, 
%the expected outcome is that the performance of the student should improve compared to a similar model trained solely on the expert-labeled data, $D_s$. 
Once trained, the student can perform end-to-end segmentation without human intervention.

\subsection{Implementation}
\begin{figure}[h!]
    \centering
    \includegraphics[width=\textwidth]{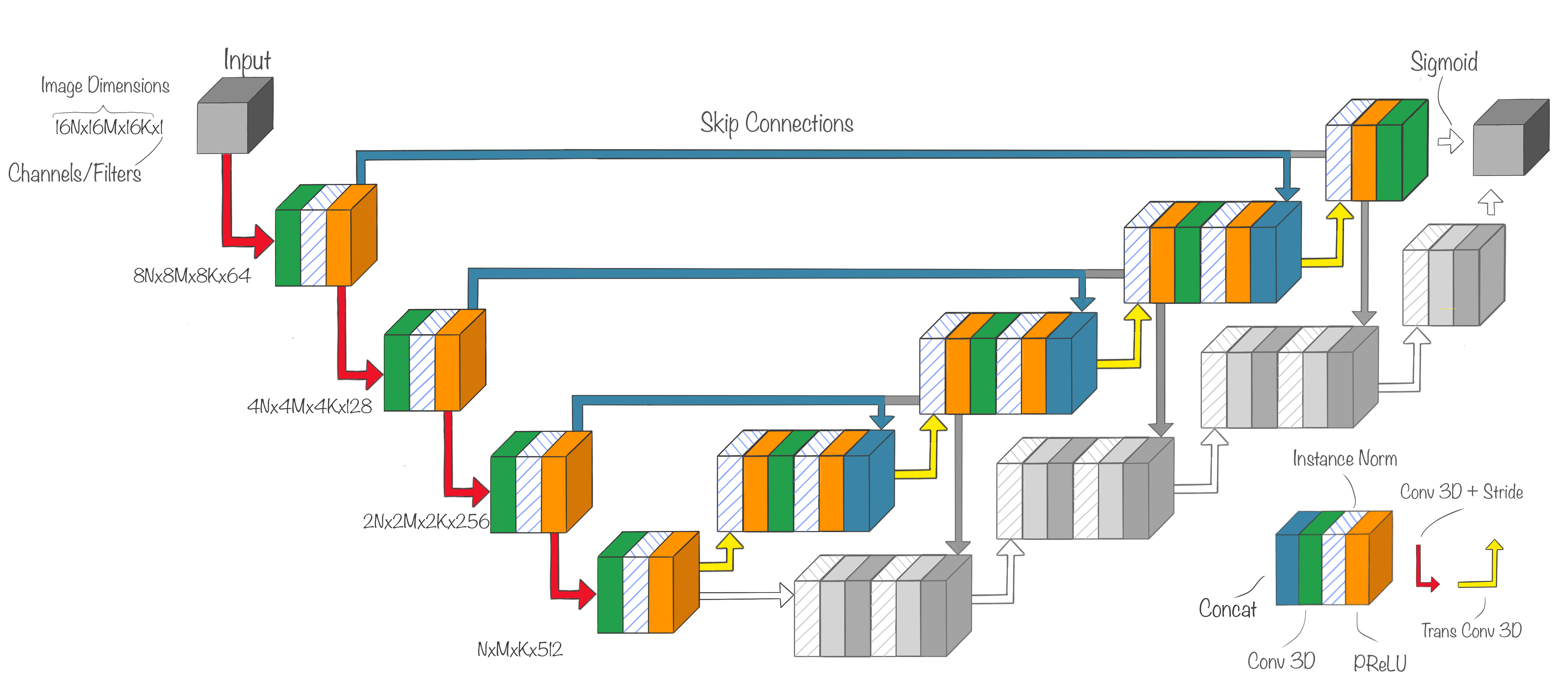}
    \caption{The network architectures. The single output (SO) Student is highlighted in colors whereas the decoder branch of the dual output (DO) Student is implicated in gray. For the DO Student the ouput of the two decoder branches are concatenated to form a dual-channeled output. The teachers have the same architecture as the SO Student, but with three downsamplings rather than four.}
    \label{fig:architecture}
\end{figure}

All our networks are based on the U-Net architecture~\cite{ronneberger2015}, and share common building blocks (see Fig. \ref{fig:architecture}). U-Net was used as it performs well as a baseline architecture, and has shown competitive performance on various datasets from different modalities, of different organs, cancer types, and data types~\cite{isensee,pedersen2021hybrid}.

The teacher consists of three levels, one of each downsampling operation, going from an image resolution of $128\times128\times128$ to $16\times16\times16$. The students are comprised of four levels. In contrast to the U-Net architecture, our design performs downsampling by applying 3D convolutions with a stride of two. We also substituted the ReLU~\cite{nair10} activation function with PReLU~\cite{He2015}. We implemented two related students: one that produces semantic segmentation output only, which we call the Single Output Student (SO Student), and one that produces an additional output approximating the bounding box surrounding the tumor in the axial plane, which we call the Dual Output Student (DO Student). The architecture of the student networks can be seen in Fig. \ref{fig:architecture}.

The Adam~\cite{kingma2017adam} optimizer with a learning rate of $10^{-4}$ was used for training until DSC validation convergence. The batch size was set to one and virtually increased to eight using accumulated gradients. The gradients were computed using the Dice Loss function~\cite{sudre2017}, based on the Dice Similarity Coefficient (DSC).

\subsection{Empirical Evaluation}
To evaluate our framework we considered two primary scenarios, each with two sub-experiments. We considered one scenario where the size of the strongly annotated dataset ($\sim 500$ images) is similar to the size of the weakly annotated dataset ($\sim 750$ images), and another scenario where the strongly annotated dataset was considerably smaller ($\sim 50$ images) than the weakly annotated dataset($\sim 1000$ images). Within each scenario, we evaluated two semi-supervised models and three fully-automatic models. Among the three fully-automatic models, one model was trained solely on strongly annotated data, whereas the two other were student networks trained both on strongly annotated data and the teacher-annotated pseudo labels.

We used different metrics for evaluating and comparing the models. The DSC was used to measure the semantic segmentation performance, whereas F1-score was used to determine object-wise localization performance. We also used DSC-TP to evaluate the segmentation accuracy considering only true positives (TPs).
%, similarly as done in a prior study~\cite{bouget2021mediastinal}.
We considered objects to be true positives if there were $\geq25\%$ overlap between the predicted mask and the GT mask, motivated by a prior study~\cite{bouget2021mediastinal}.

The test set was sampled at random and accounted for $15\%$ of the total dataset. The same split was used for all experiments to preserve fairness in evaluation. Patients with multiple scans were stratified into the three subsets: train, validation, and test. To counter the tumor size imbalance, we balanced the train and validation sets with regard to tumor sizes. Images containing tumors of more rare sizes were upsampled.

Models were trained using a workstation with a 14-core Intel Core i9 10940X @3.30 GHz CPU, 128 GB RAM, and two NVIDIA RTX 8000 (48 GB) GPUs. The most memory intensive student used, at its peak, $\sim 22.54$GB VRAM during training, but inference can be performed with 3GB VRAM. Implementation was done in Python 3.7, built upon the MONAI~\cite{monai_consortium_2020_5525502} framework (v0.4.0), using PyTorch v1.6, and CUDA 11.0. The best performing model and corresponding inference code are made openly available on GitHub as a command line tool\footnote{https://github.com/VemundFredriksen/LungTumorMask}.

\iffalse
Table \ref{tab:hardware_Specs} lists the major components of the machines used during the experiments.

\begin{table}[H]
    \centering
    \caption{Hardware Specifications}
    \vspace{5mm}
    \begin{tabular}{p{0.8in}|p{3in}}
        \textbf{Machine} & \begin{tabular}{p{0.8in}|p{2.5in}}
            \textbf{Device} & \textbf{Details} \\
        \end{tabular} \\ \hline
       Idun  & \begin{tabular}{p{0.8in}|p{2.5in}}
        GPU & Nvidia Tesla V100, 32GB\\
        CPU & Intel Xeon Gold 6148, 20 core 2.4GHz\\
       \end{tabular} \\ \hline
       Bohaga & \begin{tabular}{p{0.8in}|p{2.5in}}
        GPU & Nvidia RTX 8000, 48GB\\
        CPU & Intel Core i9 10940X, 14 core 3.3GHz\\
       \end{tabular} \\ \hline
       Malvik & \begin{tabular}{p{0.8in}|p{2.5in}}
        GPU & Nvidia Tesla V100, 32GB\\
        CPU & Intel Xeon Gold 6132, 14 core 2.6GHz\\
       \end{tabular} \\ \hline
    \end{tabular}

    \label{tab:hardware_Specs}
\end{table}
\fi

%% file: Sections/05_results.tex
\section{Results}
\subsection{Vast Strongly Annotated Dataset}
As seen in Table \ref{tab:teacher_results}, the teacher guided by the bounding boxes, outperformed the point guided (without bounding boxes as input) teacher on both datasets in terms of DSC. The difference between the two models was less prominent measured on the MSD-Lung dataset than for the NSCLC-Radiomics dataset.

\begin{table}[H]
    \centering
    \caption[Teacher Results]{Teacher Results. The best performing model with respect to mean dice similarity coefficient (DSC) is highlighted in bold.}
    \vspace{5mm}
    \begin{tabular}{p{1in}|c}
    \textbf{Dataset} &
    
    \begin{tabular}{p{1.2in}|p{1.2in}}
        \textbf{Model} & \textbf{DSC}\\
    \end{tabular} \\ \hline
    
    MSD-Lung & \begin{tabular}{p{1.2in}|p{1.2in}}
        Point Guided  & $74.78\pm11.83$ \\
        Box Guided  & $\mathbf{84.91\pm06.09}$ \\
    \end{tabular} \\ \hline
    
    NSCLC-Radiomics & \begin{tabular}{p{1.2in}|p{1.2in}}
       Point Guided  & $59.57\pm23.90$ \\
       Box Guided  & $\mathbf{86.65\pm08.77}$ \\
    \end{tabular} \\ \hline
    
    Both & \begin{tabular}{p{1.2in}|p{1.2in}}
       Point Guided  & $61.48\pm 23.29$ \\
       Box Guided  & $\mathbf{86.44\pm08.50}$ \\
    \end{tabular} \\ \hline
    \end{tabular}
    \label{tab:teacher_results}
\end{table}
\noindent For the final inference models, the DSC was highest on the MSD-Lung dataset, across all three models (see Table \ref{tab:student_results}). The best performing student network overall was the SO Student, with highest DSC on the MSD-Lung dataset. There was negligible difference between the three models on the NSCLC-Radiomics dataset. The Baseline model performed best on the MSD-Lung dataset, both in terms of DSC and F1-score.

\begin{table}[H]
    \centering
    \scriptsize
    \caption[Student Results]{Student Results. For each respective metric, the best performing models are highlighted in bold.}
    \vspace{5mm}
    \adjustbox{width=\linewidth}{
    \begin{tabular}{c|c}
    \textbf{Dataset} &
    
    \begin{tabular}{p{0.95in}|p{0.77in}p{0.72in}|p{0.7in}p{0.7in}p{0.6in}}
        \textbf{Model} & \textbf{DSC} & \textbf{DSC-TP} & \textbf{F1-score} & \textbf{Recall} & \textbf{Precision}\\
    \end{tabular} \\ \hline
    
    MSD-Lung & \begin{tabular}{p{0.95in}|p{0.77in}p{0.72in}|p{0.7in}p{0.7in}p{0.6in}}
       Baseline & $\mathbf{67.31\pm21.17}$ & $\mathbf{73.12\pm15.16}$ & $\mathbf{81.48\pm31.86}$ & $\mathbf{88.89\pm31.43}$ & $\mathbf{77.78\pm34.25}$ \\
       SO Student & $64.27\pm16.05$ & $71.32\pm8.06$ & $\mathbf{81.48\pm31.86}$ & $\mathbf{88.89\pm31.43}$ & $\mathbf{77.78\pm34.25}$ \\
       DO Student & $55.37\pm29.03$ & $70.49\pm8.82$ & $74.07\pm40.91$ & $77.78\pm41.57$ & $72.22\pm41.57$ \\
    \end{tabular} \\ \hline
    
    NSCLC-Radiomics & \begin{tabular}{p{0.95in}|p{0.77in}p{0.72in}|p{0.7in}p{0.7in}p{0.6in}}
       Baseline & $51.06\pm28.22$ & $68.81\pm18.27$ & $63.56\pm36.36$ & $\mathbf{83.82\pm36.82}$ & $56.68\pm38.65$ \\
       SO Student & $\mathbf{52.92\pm31.13}$ & $\mathbf{69.39\pm19.21}$ & $64.18\pm37.37$ & $79.90\pm39.66$ & $58.76\pm39.28$ \\
       DO Student & $52.25\pm30.18$ & $68.69\pm19.47$ & $\mathbf{68.43\pm38.71}$ & $79.17\pm39.75$ & $\mathbf{64.95\pm40.81}$ \\
    \end{tabular} \\ \hline
    
    Both & \begin{tabular}{p{0.95in}|p{0.77in}p{0.72in}|p{0.7in}p{0.7in}p{0.6in}}
       Baseline & $52.96\pm27.98$ & $69.34\pm17.97$ & $65.66\pm36.32$ & $\mathbf{84.41\pm36.27}$ & $59.14\pm38.76$ \\
       SO Student & $\mathbf{54.25\pm29.99}$ & $\mathbf{69.63\pm18.19}$ & $66.20\pm37.19$ & $80.95\pm38.90$ & $60.98\pm39.21$ \\
       DO Student & $52.61\pm30.06$ & $68.90\pm18.59$ & $\mathbf{69.09\pm39.02}$ & $79.00\pm39.97$ & $\mathbf{65.80\pm40.97}$ \\
    \end{tabular} \\ \hline
    \end{tabular}
    }
    \label{tab:student_results}
\end{table}

\subsection{Scarce Strongly Annotated Dataset}
When reducing the strongly labeled dataset, the performance of the point guided teacher was degraded, whereas the box guided teacher still performed well.

\begin{table}[H]
    \centering
    \caption[Scarcely Trained Teacher Results]{Scarcely Trained Teacher Results. The best performing model with respect to mean dice similarity coefficient (DSC) is highlighted in bold.}
    \vspace{5mm}
    \begin{tabular}{p{0.7in}|c}
    \textbf{Dataset} &
    
    \begin{tabular}{p{1.6in}|p{1.2in}}
        \textbf{Model} & \textbf{DSC}\\
    \end{tabular} \\ \hline
    
    MSD-Lung & \begin{tabular}{p{1.6in}|p{1.2in}}
       Scarce Point Guided & $48.52\pm31.18$ \\
       Scarce Box Guided  & $\mathbf{81.65\pm07.40}$ \\
    \end{tabular} \\ \hline
    
    NSCLC-Radiomics & \begin{tabular}{p{1.6in}|p{1.2in}}
       Scarce Point Guided & $43.83\pm25.65$ \\
       Scarce Box Guided  & $\mathbf{84.69\pm06.59}$ \\
    \end{tabular} \\ \hline
    
    Both & \begin{tabular}{p{1.6in}|p{1.2in}}
       Scarce Point Guided & $44.42\pm26.45$ \\
       Scarce Box Guided  & $\mathbf{84.31\pm06.77}$ \\
    \end{tabular} \\ \hline
    \end{tabular}
    \label{tab:scarce_teacher_results}
\end{table}

A similar trend applies to the final inference models (see Table \ref{tab:scarce_student_results}). The baseline model performed poorer, whereas the student networks still performed well. The same can be seen from the object-wise metrics, although the difference was more prominent. Contrary to the results shown in Table \ref{tab:student_results}, the SO Student had the highest DSC measured in this scenario. Fig. \ref{fig:scare_student_sample} shows a sample of the outputs produced by the models in the scarce scenario.

\begin{table}[H]
    \centering
    \caption[Scarcely Trained Student Results]{Scarcely Trained Student Results. For each respective metric, the best performing model is highlighted in bold.}
    \vspace{5mm}
    \scriptsize
    \adjustbox{width=\linewidth}{
    \begin{tabular}{c|c}
    \textbf{Dataset} &
    
    \begin{tabular}{p{0.95in}|p{0.77in}p{0.72in}|p{0.7in}p{0.7in}p{0.6in}}
        \textbf{Model} & \textbf{DSC} & \textbf{DSC-TP} & \textbf{F1-score} & \textbf{Recall} & \textbf{Precision}\\
    \end{tabular} \\ \hline
    
    MSD-Lung & \begin{tabular}{p{0.95in}|p{0.77in}p{0.72in}|p{0.7in}p{0.7in}p{0.6in}}
       Baseline& $26.45\pm26.56$ & $75.24\pm15.90$ & $09.10\pm13.26$ & $33.33\pm47.14$ & $05.31\pm07.80$ \\
       SO Student & $64.74\pm11.82$ & $71.56\pm10.40$ & $61.85\pm16.49$ & $\mathbf{100.0\pm0.00}$ & $47.22\pm20.79$ \\
       DO Student & $\mathbf{71.00\pm16.01}$ & $\mathbf{76.67\pm07.08}$ & $\mathbf{85.18\pm31.86}$ & $88.89\pm31.43$ & $\mathbf{83.33\pm33.33}$ \\
    \end{tabular} \\ \hline
    
    NSCLC-Radiomics & \begin{tabular}{p{0.95in}|p{0.77in}p{0.72in}|p{0.7in}p{0.7in}p{0.6in}}
       Baseline & $28.23\pm28.05$ & $55.39\pm22.80$ & $32.13\pm36.99$ & $51.47\pm49.24$ & $26.99\pm35.68$ \\
       SO Student & $51.06\pm30.75$ & $\mathbf{68.56\pm20.69}$ & $62.65\pm36.73$ & $79.41\pm39.51$ & $56.67\pm38.79$\\
       DO Student & $\mathbf{53.89\pm29.75}$ & $67.44\pm21.70$ & $\mathbf{66.96\pm35.57}$ & $\mathbf{84.56\pm35.62}$ & $\mathbf{60.44\pm38.24}$ \\
    \end{tabular} \\ \hline
    Both & \begin{tabular}{p{0.95in}|p{0.77in}p{0.72in}|p{0.7in}p{0.7in}p{0.6in}}
       Baseline& $28.02\pm27.89$ & $56.91\pm22.96$ & $29.44\pm35.82$ & $49.35\pm49.34$ & $24.45\pm34.35$ \\
       SO Student & $52.66\pm29.51$ & $\mathbf{68.98\pm19.60}$ & $62.55\pm34.98$ & $81.82\pm37.72$ & $55.56\pm37.26$ \\
       DO Student & $\mathbf{55.89\pm29.01}$ & $68.56\pm20.71$ & $\mathbf{69.09\pm35.64}$ & $\mathbf{85.06\pm35.18}$ & $\mathbf{63.12\pm38.41}$ \\
    \end{tabular} \\ \hline
    \end{tabular}
    }

    \label{tab:scarce_student_results}
\end{table}

%% file: Sections/06_discussion.tex
\section{Discussion}
In this paper, a teacher-student design to segment lung tumors from CTs has been proposed. Three datasets of two different annotation types were used for this purpose. The teacher model was first trained on the datasets that had strong annotations. It was then used to generate pseudo-strong annotations for the student. Both the teacher and the student used U-Net-like architectures, and were evaluated on segmentation performance. In addition, the student networks were evaluated on sensitivity to annotation type and sample size.

We observed that the box guided teacher outperformed the point guided teacher in both scenarios. This was expected as the bounding box annotations assist the teacher by serving as a segmentation and localization constraint. The effect of the box guidance is especially visible in the scarce scenario, where the box guided teacher achieved almost double the DSC as the point guided one. The scarce box guided teacher also outperformed the vast point guided teacher. This suggests that training a teacher on a smaller set of bounding box annotated images can be advantageous compared to training a teacher on a large set of point guided images.

Surprisingly, the students did not perform better than the baseline in the scenario with vast strongly annotated data (see Table \ref{tab:student_results}). Measured on the MSD-Lung dataset, the baseline model outperformed the two students, whereas the opposite was observed for the NSCLC-Radiomics dataset. A potential explanation might be that the Lung-PET-CT-Dx dataset contains tumors with sizes more similar to the NSCLC-Radiomics dataset than to the tumors in MSD-Lung. The introduction of the Lung-PET-CT-Dx dataset may have led the students to perform better on larger tumors, but may have degraded the results on smaller tumors typically found in MSD-Lung. Another explanation might be that the ratio between strong and weak labels were not large enough to make a noticeable difference. This was further demonstrated when the models were evaluated in the scarce scenario (see Table \ref{tab:scarce_student_results}).

In this scenario, the students significantly outperformed the baseline supervised model. This demonstrates that the introduction of suboptimal annotations into the teacher-student design can improve performance of an end-to-end segmentation model. We observed a DSC comparable with state-of-the-art performance measured on the MSD-Lung dataset, with a F1-score of $85.18$, and a DSC of $71.00$, for one of our students. Isensee \textit{et al.}~\cite{isensee} reported a DSC of $69.2$ on the MSD-Lung dataset, whereas Carvalho \textit{et al.}~\cite{carvalho} reported a DSC of $70.9$. Our model trained on only $40$ human annotated images scored marginally better, although on a different test set. The results suggest that $40$ images is not enough to train a supervised model, but enough to train a semi-supervised model that can enhance a supervised model by increasing the available data in a cheaper way than manual delineation.

\begin{figure}[h!]
    \centering
    \includegraphics[width=\textwidth]{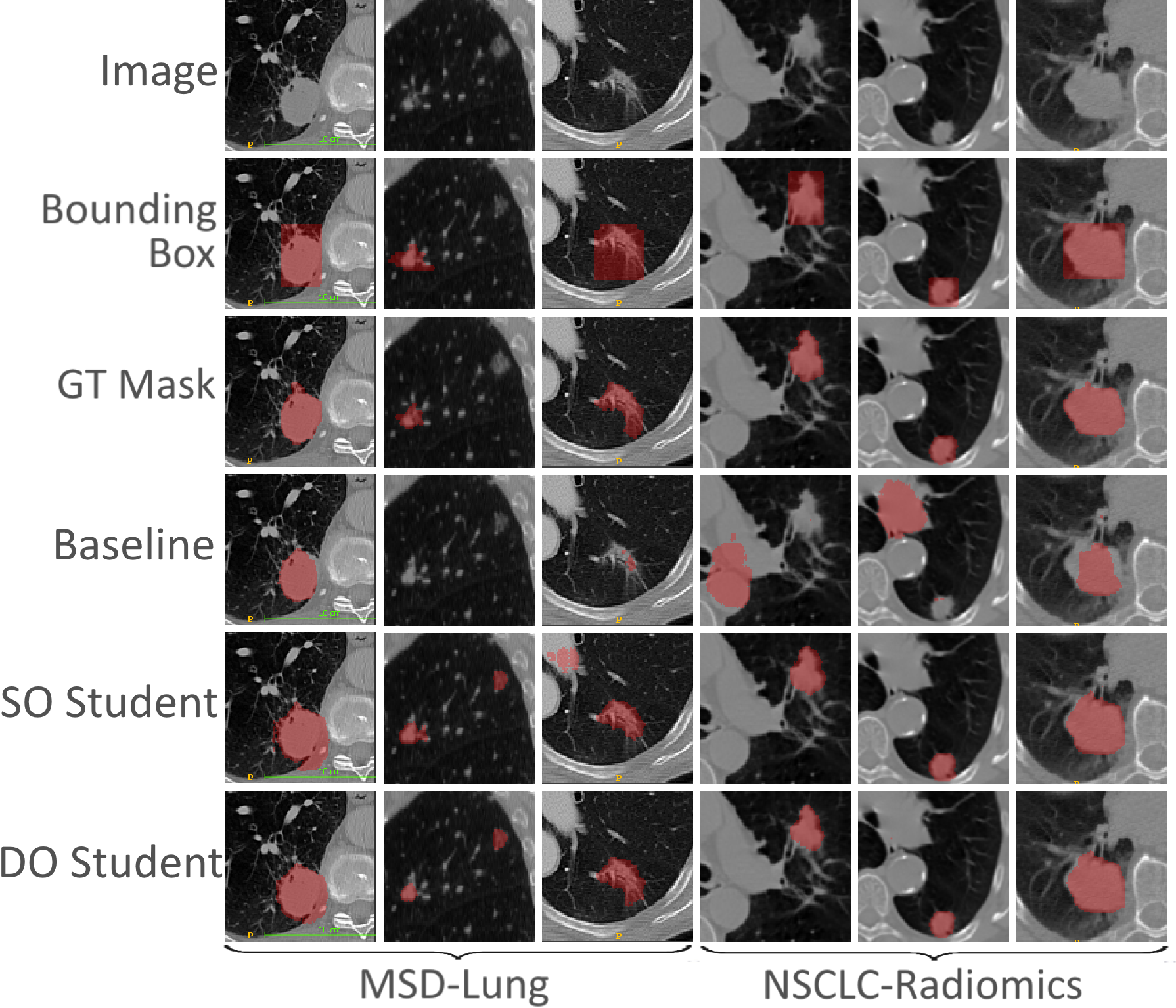}
    \caption{A sample of the results produced by the scarce students on the test set. The figure shows the input image, bounding box, and ground truth (GT) mask in the three top rows, respectively. The baseline model, single output (SO) Student, and dual output (DO) Students corresponding outputs are shown in the three bottom rows, respectively.}
    \label{fig:scare_student_sample}
\end{figure}

The datasets were of varying quality. The MSD-Lung dataset was of high standard, whereas the NSCLC-Radiomics dataset was less so. Other publications that used the NSCLC-Radiomics dataset reported heavy data sanitation, effectively removing large parts of the dataset~\cite{pang,kamal}. We did not override the expert's annotations, as we also seek to handle suboptimal annotations, if these should be present in a data set. The flawed dataset explains why the difference between the box guided and point guided teacher is larger on the NSCLC-Radiomics dataset than MSD-Lung. Images where the tumor is poorly, or even completely wrongly annotated, the box guided teacher can rely on the bounding boxes to achieve a good DSC, but since the annotation itself is wrong, the point guided teacher struggles.

\subsection{Limitations}
One of the major limitations in this experiment was the scarce amount of data. The test set was sampled randomly from each dataset. It is plausible that a different sample of the test set would have given different result. Although K-fold cross validation could be used to eliminate this concern, it was dropped due to time limitations. K-fold cross validation is a time consuming strategy. It depends on training K different models, which would take a considerable amount of time, even with a small K in our situation. Since our method is a two step method that relies on two training steps, the K-fold cross validation would take nearly double the time of a similar single-step method as well. 

Another limitations of this experiments was that the students were sensitive to voxel spacing. By reducing the voxel spacing during normalization/preprocessing, thus increasing the resolution of the image, the DSC did not improve, but actually degraded. Therefore, it is possible that the proposed architecture is sensitive to small adjustments in the preprocessing pipeline.

\subsection{Future work}
An alternative to using 2D bounding boxes in the axial plane, is to use a 3D bounding box. As one 3D bounding box contains much fewer corners than multiple 2D bounding boxes, this could further reduce annotation load. It is reason to believe that a teacher trained on 3D boxes will perform worse than one trained on 2D axial boxes. However, if the reduction in annotation load is significant, the amount of data that can be annotated for the teacher might weigh up the loss in precision of the annotation. After all, this is the very fundamental idea behind the teacher-student design. However, we feel that a much larger dataset should be used to explore this properly.

The main motivation of using a teacher-student design is to improve models by learning from additional suboptimal annotated or unannotated data. We observed a benefit of using such a design for lung tumor segmentation in CTs. However, a single-step teacher might not be sufficient. It has been proposed to train both the teacher and student end-to-end in an iterative fashion~\cite{caron2021emerging}. This makes sense as the teacher could improve from the student's feedback. Especially from multiple students, which iteratively could improve the students as well, as the teacher become more experienced. However, for 3D applications this is likely infeasible. Alternatively, one could potentially use multiple teachers, trained on different types of images that focus on different lung tumor types and sizes. Having specialized teachers to train the student in an ensemble manner makes sense as it more closely represents the natural teacher-student relation from academia.

%% file: Sections/07_conclusion.tex
\section{Conclusion}
We present the first known implementation of a mixed-supervised teacher-student framework for lung tumor segmentation from CT images. Our method utilized both semantic and axial bounding box annotations to maximize lung tumor segmentation performance. We demonstrated that with sufficient bounding box annotated data, our teacher-student framework achieved state-of-the-art performance, even with scarce semantic annotated data.

%% file: Sections/08_acknowledgements.tex
\subsection*{Acknowledgements}
This work was conducted at the machines Idun~\cite{sjalander2019}, Malvik\footnote{A cluster node owned and maintained by NTNU AI Lab}, and Bohaga\footnote{A machine owned by NTNU, maintained by Gabriel Kiss}.

\subsection*{Author contributions}
VF and SS were involved in code writing;
AP and FL supervised the implementation; Computational resources were provided by GK and FL; FL and TL contributed to funding acquisition; TL and AP provided domain-level expertise; All authors contributed to writing-original draft preparation and to writing-review and editing. All authors read and approved the final manuscript.

\subsection*{Funding}
The project was supported by The Norwegian National Advisory Unit for Ultrasound and Image-Guided therapy at St. Olavs hospital; The Liaison Committee for Education, Research and Innovation in Central Norway [grant number 2018/42794];
The Cancer Foundation, St. Olavs hospital, Trondheim University Hospital [grant number 13/2021];
SINTEF; The EEA Grant project "Improving Cancer Diagnostics in Flexible Endoscopy Using Artificial Intelligence and Medical Robotics (IDEAR)" [grant number 19/2020];
Open Access funding was provided by the Norwegian University of Science and Technology, Trondheim, Norway.

%% file: Sections/09_declarations.tex
\section*{Declaration}

\subsection*{Conflicts of interest}
The authors have nothing to disclose.

\subsection*{Data availability}
Only open datasets were used to conduct the experiments, and are therefore available to reproduce the experiments.

\subsection*{Code availability}
A command line tool for end-to-end lung tumor segmentation of our best performing model is made openly available on GitHub: \url{https://github.com/VemundFredriksen/LungTumorMask}.